\documentclass[journal=jacsat,manuscript=article]{achemso}

\usepackage{chemformula} 
\usepackage[T1]{fontenc} 
\usepackage[utf8]{inputenc}

\newcommand{\fcal}{ \mathcal{F} }
\newcommand{\rvec}{ \mathbf{r} }
\newcommand{\Rvec}{ \mathbf{R} }
\newcommand{\fvec}{ \mathbf{f} }

\newcommand{\uhat}{\hat{\textbf{u}}}
\newcommand{\rhat}{\hat{\textbf{r}}}

\newcommand{\dcal}{ \mathcal{D} }

\DeclareUnicodeCharacter{2212}{-}
\DeclareUnicodeCharacter{2032}{'}

\author{Javier Diaz }
\affiliation{School of Mathematics and Physics, University of Lincoln. Brayford Pool, Lincoln, LN6 7TS, UK}
\author{Marco Pinna}
\affiliation{School of Mathematics and Physics, University of Lincoln. Brayford Pool, Lincoln, LN6 7TS, UK}
\email{mpinna@lincoln.ac.uk}
\author{Andrei V. Zvelindovsky}
\affiliation{School of Mathematics and Physics, University of Lincoln. Brayford Pool, Lincoln, LN6 7TS, UK}
\author{Ignacio Pagonabarraga}
\affiliation{Departament de Fisica de la Materia Condensada, Universitat de Barcelona, Marti i Franques 1, 08028 Barcelona, Spain }
\affiliation{CECAM, Centre Europ\'een de Calcul Atomique et Mol\'eculaire, \'Ecole Polytechnique F\'ed\'erale de Lausanne,
Batochime - Avenue Forel 2, 1015 Lausanne, Switzerland }
\affiliation{Universitat de Barcelona Institute of Complex Systems (UBICS), Universitat de Barcelona, 08028 Barcelona, Spain}
\author{Roy Shenhar}
\affiliation{Institute of Chemistry,
The Hebrew University of Jerusalem,
Edmond J. Safra Campus,
Givat Ram, Jerusalem, Israel 9190401}
\email{roys@huji.ac.il}

\title[An \textsf{achemso} demo]
  { 
  Block copolymer-nanorod co-assembly in thin films:
  effects of rod-rod interaction and confinement	  
  }

\abbreviations{*********************}
\setkeys{acs}{keywords = true}

\begin{document}
\maketitle
\begin{center}
\includegraphics[height=3.5cm]{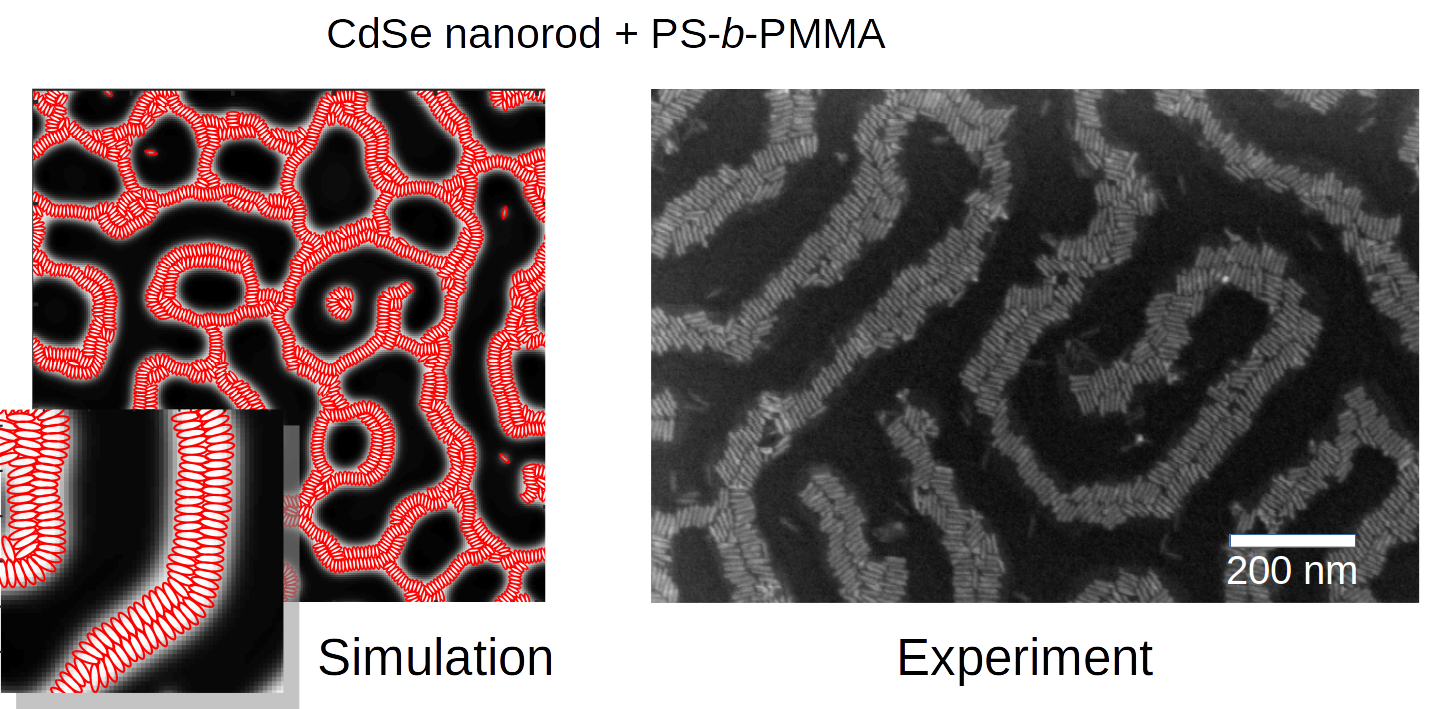}
\end{center}
\begin{center}
For Table of Contents use only
\end{center}

\begin{abstract}
Simulations and experiments of nanorods (NRs) show that co-assembly with block copolymer (BCP) melts leads to the formation of a superstructure of side-to-side NRs perpendicular to the lamellar axis. A mesoscopic model is validated against  scanning electron microscopy (SEM) images of CdSe NRs mixed with polystyrene-\textit{block}-poly(methyl methacrylate). It is then used to study the co-assembly of anisotropic nanoparticles (NPs) with a length in the same order of magnitude as the lamellar spacing. The phase diagram of BCP/NP is explored as well as the time evolution of the NR. NRs that are slightly larger than the lamellar spacing are found to rotate and organise side-to-side with a tilted orientation with respect to the interface. 
Strongly interacting NPs are found to dominate the co-assembly while weakly interacting nanoparticles are less prone to form aggregates and tend to form well-ordered configurations. 
\end{abstract}



Block copolymer (BCP) melts can self assemble into well-ordered mesophases\cite{bates_block_1999,bates_block_1990,bates_fluctuations_1994,matsen_unifying_1996}, which are repeated periodically with a domain size $H_0$, typically of the order of $1-300$ nm \cite{hashimoto_domain-boundary_1980}. 
This periodicity makes BCPs excellent matrices to host nanoparticles (NPs), which can be localised in specific regions of the phase-separated BCP\cite{okumura_nanohybrids_2000,kim_effect_2006}. 

Mixtures of BCPs and colloids have long been studied using  theory\cite{pryamitsyn_strong_2006,pryamitsyn_origins_2006}
simulations
\cite{huh_thermodynamic_2000,thompson_block_2002,thompson_predicting_2001} 
and  experiments\cite{bockstaller_size-selective_2003,bockstaller_block_2005}
due to the interesting behaviour resulting from the co-assembly of selective nanoparticles and phase-separated block copolymer. 
Nanorods (NR) have attracted considerable attention as  constituents of functional polymer nanocomposite materials\cite{hore_functional_2014}.  
The orientational degree of freedom of anisotropic colloids introduces new possibilities of BCP/NP co-assembly,  thanks to the intrinsic ordered structures of the neat BCP (lamellar, cylindrical, etc). 
 For instance, gold NRs have been found to orient along the lamellar domain axis when confined in one of the symmetrical phases\cite{deshmukh_two-dimensional_2007,tang_self-assembly_2009}. 
 Similarly, gold NRs template the direction of the cylindrical domains in an asymmetrical diblock copolymer mixture\cite{laicer_gold_2005}. 
 Ordered arrays of aligned NRs were achieved by Thorkelsson \cite{thorkelsson_direct_2012,thorkelsson_end--end_2013} in the co-assembly of BCP and anisotropic particles, where NRs were organised in an end-to-end configuration. 
Nanoplates alignment in a lamellar-forming BCP has been recently studied\cite{krook_alignment_2018}. 

Experiments have reported the existence of an ordered phase when NRs are mixed with asymmetric diblock copolymer \cite{ploshnik_co-assembly_2010,ploshnik_hierarchical_2010} in thin films. 
Shenhar and Banin studied polystyrene-\textit{block}-poly(methyl methacrylate) (PS-\textit{b}-PMMA) copolymers mixed with PS-modified CdSe NRs, and found that NRs preferentially organized in a side-to-side configuration, forming long rows of particles in the PS domains, with an orientation normal to the interface between BCP domains, ie, perpendicular to the direction of the lamella domain. 
Furthermore, the number of rows and degree of order could be related to the size of the NRs and copolymer spacing.

Theoretical and computational works have studied the self-assembly of BCP and anisotropic NPs. 
Dissipative Particle Dynamics (DPD) has been largely used, thanks to the ability to combine several beads into rod-like sequences. 
Zhang et al studied the phase behavior of such systems and the orientation of nanoparticles \cite{he_effect_2009,he_phase_2009,he_mono-_2010}, and  the effect of shear in the global orientation has also been quantified \cite{pan_dynamic_2011}. 
Osipov et al \cite{osipov_spatial_2016,osipov_induced_2017,osipov_phase_2018}
used strong and weak segregation  theory to determine the distribution of anisotropic particles in a diblock copolymer, with a low fraction of NPs present in the system.


Here, we make use of the considerably fast Cell Dynamic Simulation (CDS) method to simulate the BCP dynamics while Brownian Dynamics describes the assembly of ellipsoidal colloids. These simulations are compared with experiments involving CdSe NRs, in order to study the co-assembly of colloids within BCP domains. Simulations are used to gain insight over the behavior and occurrence of the orientational order of anisotropic colloids. Ellipses are used to mimic the shape of NRs.

The Cell Dynamic Simulation method has been used extensively both in pure BCP systems \cite{ren_cell_2001,pinna_large_2012,pinna_diblock_2011,dessi_cell_2013} and nanocomposite systems\cite{pinna_modeling_2011}, reproducing experiments such as aggregation of incompatible colloids \cite{diaz_cell_2017,ploshnik_hierarchical_2013}
and NP-induced phase transitions
\cite{diaz_phase_2018}. 
Its relative computational speed makes it suitable to study properties that involve large systems over extended times, while the phenomenological approach in its model limits its validity in the microscopic realm. 
This hybrid method permits to explore the high NP filling fraction regime, in which the presence of NPs introduces considerable perturbations to  the neat BCP matrix, such as  morphological phase transitions. 

We aim to systematically study the phase behaviour of a polymer composite system made of diblock copolymer  and anisotropic NPs, restricting to the case of NPs which are compatible with one of the copolymers. 
Size, shape and number of particles were explored, to address its effect on both the diblock copolymer morphology and especially the colloidal assembly. Several length scales are present in polymer nanocomposite systems\cite{langner_mesoscale_2012}, specially in the case of NRs or elliptical particles. This variety of sizes has been shown to result in interesting effects of confinement, and presents a challenge for its study. 
Simulations are compared with experimental results, to first assess its validity and then explore several parameters and configurations which are experimentally more challenging. 

\section{Model}

The evolution of the BCP/colloids  system is determined by the excess free energy which can be separated as 
\begin{equation}
\fcal _{tot} = \fcal_{pol}+\fcal_{cc} +\fcal_{cpl}
\end{equation}
with $\fcal_{pol}$ being the free energy functional of the BCP melt, $\fcal_{cc}$ the colloid-colloid interaction and the last contribution being the coupling term between the BCP melt and the colloids. 


The diblock copolymer is characterized by the order parameter $\psi ( \rvec ,t   )$ which represents the differences in the local volume fraction for the copolymer A and B 
\begin{equation}
\psi (\rvec,t )= 
\phi_A (\rvec,t)
-
\phi_B (\rvec,t)
+(1-2f_0)
\end{equation}
with respect to the relative volume fraction of A monomers in the diblock, $f_0= N_A/ (N_A +N_B)$.
The order parameter must follow the continuity equation in order to satisfy the mass conservation of the polymer: 
\begin{equation}
\frac{\partial\psi ( \rvec, t )}{\partial t}=
-\nabla\cdot \mathbf{j} (\rvec ,t ) 
\end{equation} 

If the polymer relaxes diffusely towards equilibrium, the order parameter flux can be expressed in the form 
\begin{equation}
\mathbf{j } (\rvec,t    )=
-M \ \nabla \mu (\rvec , t )
\end{equation}
as a linear function of the order parameter chemical potential
\begin{equation}
\mu (\rvec , t )=
\frac{\delta \fcal_{tot}  [ \psi] }{ \delta \psi}
\end{equation}

Introducing these equations into the continuity equation and taking into account the thermal fluctuations we obtain the Cahn-Hilliard-Cook equation (CHC)
\begin{equation}
\frac{\partial\psi ( \rvec, t )}{\partial t}=
M\ \nabla^2 \left[
\frac{\delta \fcal_{tot}  [ \psi] }{ \delta \psi}\right]
+
\xi ( \rvec, t)
\label{eq:ellipse.cahn}
\end{equation}
where $M$ is a phenomenological  mobility constant and $\xi$ is a white Gaussian random noise which satisfies the fluctuation-dissipation theorem\cite{ball_spinodal_1990}. 

The copolymer free energy is a functional of the local order parameter which can be expressed in terms of the thermal energy $k_B T$ as
\begin{equation}
 \fcal_{pol}  [ \psi (\rvec ) ]=
\int d\rvec \left[
H(\psi) +\frac{1}{2} D | \nabla\psi  |^2   
\right]
+ \\
\frac{1}{2} B \int d\rvec  \int d\rvec' \ 
G(\rvec -\rvec ' )\psi(\rvec)\psi(\rvec') 
 \end{equation}
where the first and second terms are the short and the long-range interaction terms respectively, the coefficient $D$ is a positive constant that accounts for the cost of local polymer concentration inhomogeneities, the Green function $G(\rvec-\rvec' )$ for the laplace Equation satisfies $\nabla^2 G(\rvec-\rvec') = -\delta (\rvec-\rvec')$, $B$  is a parameter that introduces a chain-length dependence to the free energy\cite{hamley_cell_2000}. 
The lamellar periodicity is $H_0 \propto 1/\sqrt{B}$.  The local free energy  is\cite{hamley_cell_2000,ren_cell_2001}, 
\begin{equation}
  H(\psi )  = 
 \frac{1}{2}\left[   
 -\tau_0+ A(1-2f_0)^2
 \right]   \psi ^2  \\
 +\frac{1}{3} v (1-2f_0)\psi^3 
 +\frac{1}{4}u \psi^4
 \label{eq:Hpsi}
 \end{equation}
where $\tau_0,A,v,u $ are phenomenological parameters\cite{ren_cell_2001} which can be related to the block-copolymer molecular specificity. Previous works\cite{pinna_modeling_2011,ren_cell_2001,ohta_equilibrium_1986} describe the connection of these effective parameters to the BCP molecular composition.  $\tau ' = -\tau_0+A(1-2f_0)^2$, $D$ and $B$ can be expressed\cite{ohta_equilibrium_1986} in terms of degree of polymerization $N$, the segment length $b$  and the Flory-Huggins parameter $\chi$(inversely proportional to temperature)  . 
Subsequently, we will consider $u$ and $v$  constants\cite{leibler_theory_1980}, which define all the parameters identifying the BCP local  free energy $H(\psi)$ . As  previously shown \cite{sevink_selective_2011,pinna_mechanisms_2009}, CDS can be used along with more detailed approaches like dynamics self-consistent field theory (DSCFT), using CDS as a precursor in exploring parameter space due to the computationally inexpensiveness nature of CDS. We can express the time evolution of $\psi$ , Equation \ref{eq:ellipse.cahn}, using CDS as 
\begin{equation}  
 \psi ( \rvec_i , t+1   )= \psi (\rvec_i,t )-  
\delta t [ 
\langle \langle \Gamma (\rvec_i, t \rangle\rangle \\
- \Gamma (\rvec_i, t ) +  
B     [ 1- P (\rvec_i, t) \psi (\rvec_i,t )]   -\eta \xi (\rvec_i,  t)       ] 
]
 \label{eq:ellipse.time_evol}
\end{equation}
$\rvec_i$ being the position of the node $i$ at a time $t\delta t$, and the isotropic discrete  laplacian for a quantity $X$ is given by \cite{oono_study_1988}
$\frac{1}{\delta x^2}  [ \langle\langle X \rangle\rangle -X  ] $. Specifically, we  will use 
\begin{equation}
\langle \langle \psi \rangle \rangle = \frac{1}{6}  \sum_{NN}  \psi   +\frac{1}{12} \sum _{NNN} \psi
\end{equation} 
NN, NNN meaning nearest neighbours and next-nearest neighbours, respectively, for the two dimensional case.
The lattice spacing is $\delta x $.

In Equation \ref{eq:ellipse.time_evol} we have introduced the auxiliary function 
\begin{equation}
\Gamma (\rvec, t ) =
g( \psi (\rvec, t)  )- \psi (\rvec, t)+
D \left[ 
\langle\langle   \psi (\rvec, t)    \rangle \rangle   -\psi (\rvec, t)
\right]
\end{equation}
and also, the map function \cite{bahiana_cell_1990,ren_cell_2001}
\begin{equation}
g (\psi)= -\tau ' \psi -v (1-2f_0)\psi^2 -u \psi^3
\end{equation}


\subsection{Polymer/colloid interaction }

Contrary to the polymeric matrix, a suspension of $N_p$  nanoparticles describes each colloidal NP individually through the center of mass and orientation degrees of freedom $\Rvec_i,\phi_i$. 
The interaction between the polymer and colloids is introduced through a contribution to the free energy $\fcal_{cpl}$, which must take into account the fact that colloids may have a preference for the A-block of the A-\textit{b}-B BCP. 
The simplest free energy that satisfies that is 
\begin{equation}
\fcal_{cpl} =
\sum _{i=1}^{N_p}
\sigma \int d \rvec\ \psi_{c } (\rvec,\Rvec_i,\phi_i) \left[ \psi (\rvec )-\psi_0   \right]^2
\label{eq:ellipse.coupling}
\end{equation}
where $\sigma$ defines the strength of the interaction between polymer and colloids, and $\psi_0$ describes the affinity of NPs with  the BCP. 

In previous works\cite{pinna_modeling_2011}, the size, shape and core/shell properties of the NP are described through the tagged function\cite{tanaka_simulation_2000} $\psi_c(\rvec)$. 
In order to account for non-spherical colloids, we generalise the spherical shape into an non-rotated ellipse placed at $\Rvec_i=(0,0)$ as 
\begin{equation}
    \psi_{c}(x,y)=\exp\left[ 
    1-\frac{1}{1-
    \left(\frac{x}{a}\right)^2-\left(\frac{y}{b}\right)^2}
    \right] 
    \label{eq:ellipse.psic}
\end{equation}
which can be trivially extended for an arbitrary rotation $\phi_i$. 
The particle shape and size is characterised by a major semiaxis $a$, and hard-core major semiaxis $a_0=a/\sqrt{1+1/\ln 2}$, and the same relationship holds for the minor semiaxis $b$. 
The ratio $e=b/a$ accounts for anisotropy of the ellipse. 
The tagged function is $\psi_c(\rvec) = 0$ outside of the ellipsoids, that is, for $(x/a)^2+(y/b)^2>1$.

\subsection{Interparticle potential}
In order to introduce colloid-colloid interactions we require an orientational-dependent pairwise additive potential. 
The potential we use is the standard Gay-Berne (GB) potential \cite{gay_modification_1981,berne_gaussian_1972} which derives from a Gaussian overlap study of ellipsoids, making it suitable to our interactions. 
The GB potential has been widely used to describe liquid crystals\cite{de_miguel_liquid_1991,berardi_monte_1993}. 
The interparticle potential can be written as 
\begin{equation}
 V(\uhat_1,\uhat_2,\rvec)=
\epsilon(\uhat_1,\uhat_2,\rhat) \\
\left[
\left(
\frac{1}{r-\sigma(\uhat_1,\uhat_2,\rvec)}
\right)^{12}
-  
\left(
\frac{1}{r-\sigma(\uhat_1,\uhat_2,\rvec)}
\right)^{6}
\right]
 \end{equation}
which is a modified Lennard-Jones interaction with anisotropic length and energy scales, $\sigma(\uhat_1,\uhat_2,\rhat)$ and $\epsilon(\uhat_1,\uhat_2,\rhat)$, respectively. 
The centre-to-centre distance is $r$ while $\uhat_i$ stands for the orientation of the major axis of particle $i$.   
This potential provides a length scale that describes the anisotropy of the ellipsoid
\begin{equation}
 \sigma(\uhat_1,\uhat_2,\rhat)=
2b \\
 \left\lbrace
1- 
\frac{1}{2}\chi 
\left[
\frac{(\rvec\cdot \uhat_1+\rvec+\cdot\uhat_2)^2}{1+\chi(\uhat_1\cdot\uhat_2)}
+
\frac{(\rvec\cdot \uhat_1-\rvec+\cdot\uhat_2)^2}{1-\chi(\uhat_1\cdot\uhat_2)}
\right]
\right\rbrace^{-1/2}
 \end{equation}
  and takes a value $2b$ at the side-to-side configuration. The energetic anisotropy is described with two parameters: $U_0$ describes the strength of the interaction while $\epsilon_r=\frac{\epsilon_e}{\epsilon_s}$ is an expression of the anisotropy of the wells. The depth of the well is given by 
  \begin{equation}
  \epsilon(\uhat_1,\uhat_2,\rhat)=
  \epsilon(\uhat_1,\uhat_2)
  \epsilon'^2(\uhat_1,\uhat_2,\rhat)
  \end{equation}
with 
\begin{equation}
\epsilon(\uhat_1,\uhat_2)=
U_0 
\left[
1-\chi^2(\uhat_1\cdot \uhat_2)^2)
\right]^{-1/2}
\end{equation}
and 
\begin{equation}
\epsilon'(\uhat_1,\uhat_2,\rhat)=
1-\frac{1}{2}\chi' 
\left[
\frac{(\rvec\cdot \uhat_1+\rvec+\cdot\uhat_2)^2}{1+\chi'(\uhat_1\cdot\uhat_2)}
+
\frac{(\rvec\cdot \uhat_1-\rvec+\cdot\uhat_2)^2}{1-\chi'(\uhat_1\cdot\uhat_2)}
\right]
\end{equation}
where two anisotropy parameters are introduced, regarding length and energy, respectively,
\begin{equation}
\chi=\frac{a^2-b^2}{a^2+b^2}; \ 
\chi'=\frac{\epsilon_s^{1/2}-\epsilon_e^{1/2}}{\epsilon_s^{1/2}+\epsilon_e^{1/2}}
\end{equation}

\subsection{Colloid Dynamics: Brownian Dynamics } 

Since the NPs are anisotropic, the equation of motion  does not involve only the friction constants but a diffusion tensor, $\dcal$. In general\cite{han_brownian_2006}, 
\begin{equation}
\frac{d \rvec}{dt }=
\dcal_t \cdot \fvec
\end{equation}
while the particle's orientational degree of freedom relates to the random ($M_r$) and exerted torques as
\begin{equation}
\frac{\partial \phi_i}{\partial t }=
\left( M_i+M_r \right)/\gamma_\phi; \ \ M_i= -\frac{\partial \fcal }{\partial \phi_i}
\end{equation}
with \cite{hagen_brownian_2011}
\begin{equation}
\dcal_t= \bar{\dcal}  \mathcal{I}+\frac{1}{2} \Delta \dcal 
\begin{pmatrix} \cos 2\phi & \sin 2\phi \\ \sin 2\phi  & -\cos 2\phi \end{pmatrix}
\end{equation}
and $\bar{\dcal}=\frac{1}{2} (D_a+D_b)$ and $\Delta \dcal = D_a-D_b$, $D_a$ and $D_b$ being the diffusion constants along each axis. The values of $D_a,D_b$ and $\gamma_{\phi}$ are derived from the expressions obtained by Perrin
\cite{zheng_self-diffusion_2010,happel_low_1983,perrin_mouvement_1934}

\subsection{Order parameter}
To describe the orientation of the anisotropic NP, an order parameter can be used, which has been extensively employed in nematic liquid crystal systems,  
\begin{equation}
S=\langle
2(\uhat\cdot \mathbf{P})^2 -1
\rangle
\label{eq:ellipse.S}
\end{equation}
which is an average over all particles of the scalar product of the orientation unit vector $\uhat$ and a local unit vector $\mathbf{P}$ that is related to the gradient of the polymer order parameter $\psi(\rvec,t)$. This unit vector $\mathbf{P}$ is normal to the interface between copolymer domains. 


\section{Results and discussion}

As a first approach, we study the condition for the appearance of an ordered phase in a BCP with different compositions   $f_0$, which gives rise to a variety of BCP morphologies.  
After that, the role of the NR length will be described, in relation to the BCP periodicity. 
Finally, the role of the NP-NP interaction is asserted taking into account several initial conditions. 


We introduce dimensionless parameters rescaling $D\to D/\delta x ^2$ and $B\to B \delta x ^2$. 
Lengths are expressed in terms of grid points. 
The standard values of CDS\cite{ren_cell_2001,pinna_large_2012,pinna_modeling_2011} will be used 
$
\tau_0=0.35, u=0.5,v=1.5,A=1.5,D=1.0
$
while the BCP/NP interaction is set to $\sigma=1.0$. A cell spacing $\delta x =0.5$ and time discretisation $\delta t=0.1$ are chosen. 
Unless otherwise specified, the NP size is set to $a_0=2$ and $e=0.3$ while the BCP periodicity is determined by the CDS parameter $B=0.002$. 
The NP thermal energy is set to $k_BT=0.1$.  
The box size of simulations is $128\times 128$ except for larger systems which are explicitly stated in the text.

This work focuses on A-block compatible NPs inspired by experimental  ordered hierarchical structures of NRs in BCP\cite{ploshnik_co-assembly_2010,ploshnik_hierarchical_2010}, by selecting a value of the affinity $\psi_0=-1$ in reduced units with  the equilibrium value of $\psi$.   
The anisotropy of experimental NRs is modelled with ellipsoidal NPs with a Gay-Berne potential. 
Recently, a generalised approach to NP shape has been presented to simulate superellipses immersed in BCP \cite{diaz_nonspherical_2019}, including rectangular-shaped NPs. 
Nonetheless, the lack of an appropriate NP-NP potential limits the realistic comparison with experiments. 
We expect that, despite the differences in shape, the inclusion of anisotropic shape and orientation-dependent NP-NP potential will be sufficient to mimic experimental results, while limiting the possibility to establish a one-to-one comparison between experiments and simulations.

\subsection{Phase diagram of A-compatible ellipsoidal colloids}

The phase diagram of diblock copolymer/colloids has been widely studied both for nanospheres\cite{huh_thermodynamic_2000}  and anisotropic NPs\cite{tang_self-assembly_2009}. 
The presence of NPs which are compatible with one of the blocks  increases the effective overall volume fraction of the hosting domain, which in turn results in a phase transition. 
As a first approach to a system of BCP and anisostropic NPs, we explore  the effect that ellipse-shaped colloids have on the BCP morphology, by analysing the phase of BCP with arbitrary composition $f_0$ in the presence of a filling fraction $\phi_p$ of  NPs.

\begin{figure}[hbtp]
\centering
\includegraphics[width=0.75\linewidth]{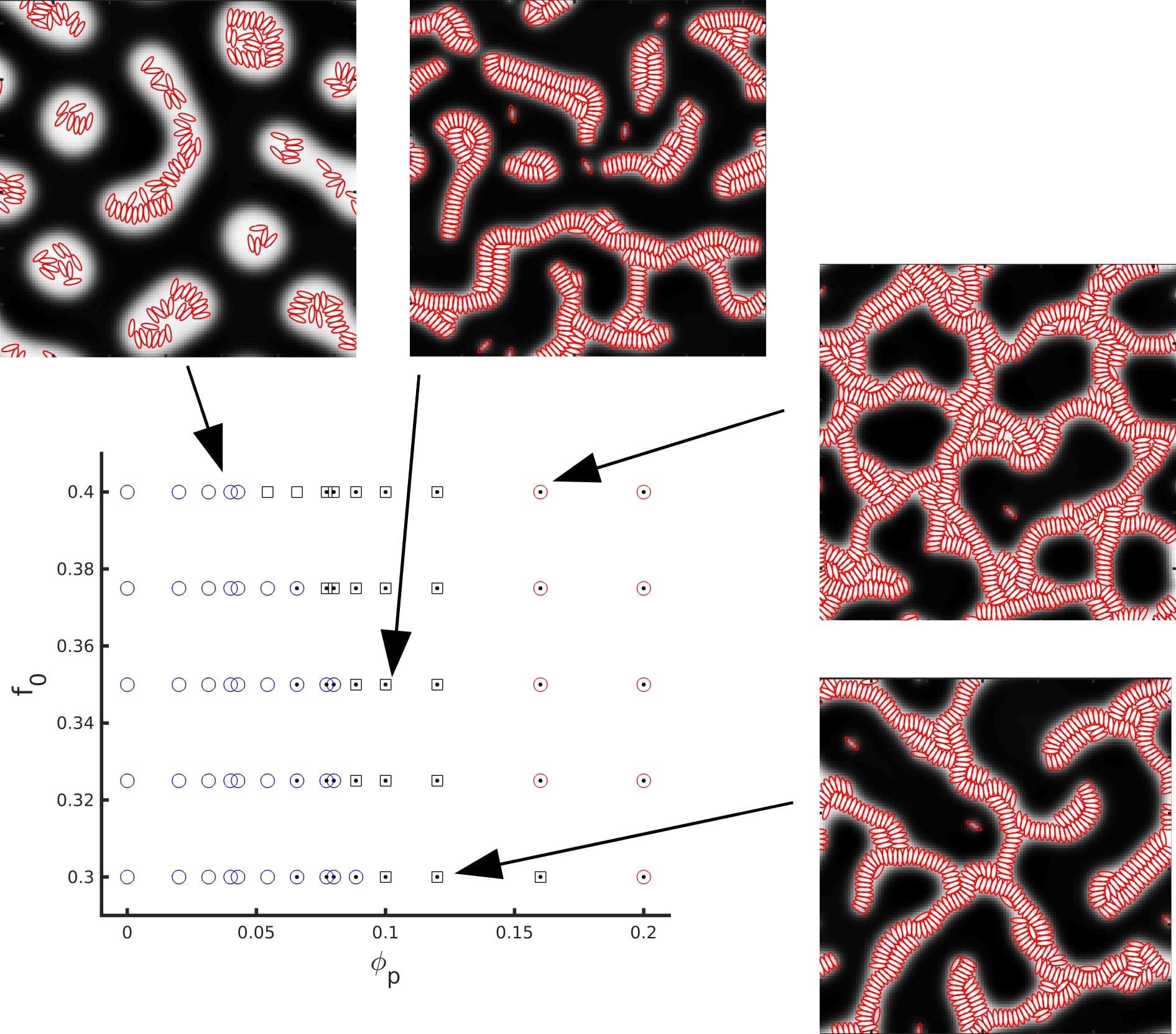}
\caption{Phase diagram of a diblock copolymer nanocomposite system characterised by a filling fraction $\phi_p$ of ellipsoidal colloids and $f_0$ volume fraction of the A blocks in the neat BCP.
 Squares, blue circles and red circles stand for lamellar, cylindrical phase and inverted cylindrical phase, respectively. Dotted markers represent phase points in which $S>0.3$ (eq. \ref{eq:ellipse.S}) ie, where ellipsoids are aligned mostly normal to the interface. }
\label{fig:ellipse.phase_diagram}
\end{figure}

In Figure \ref{fig:ellipse.phase_diagram} the filling fraction of ellipsoids is explored for different BCP compositions, $f_0$. The NP-NP interaction scale is set to $U_0/\sigma = 0.01$ so that the interparticle potential is not dominating over the BCP-NP interaction.  
As expected, at low filling fraction , $\phi_p$, particles are simply segregated within  their preferred phase (blue) which within the tested range of $f_0$ is the minority phase. 
In the absence of constrains  by the BCP (ie. a low local filling fraction) ellipsoids display no orientational order.
Furthermore, the BCP maintains a cylindrical phase (circular domains, in two dimensions).

At higher filling fractions, the ellipsoids enlarge the hosting domains to a point in which a cylinder-to-lamellae phase transition is induced. 
At the same time, higher filling fractions lead to a particular ordered phase in the colloids: ellipsoids prefer to orient normal to the interface and with a side-to-side interparticle configuration. 
This phase has been reported experimentally by Shenhar and Banin\cite{ploshnik_co-assembly_2010,ploshnik_hierarchical_2010}, where ordering was reported to be driven by both attractive NP-NP interaction and minimisation of the repulsive interactions between the NRs and the B phase. 

\begin{figure}[hbtp]
\centering
\includegraphics[width=0.75\linewidth]{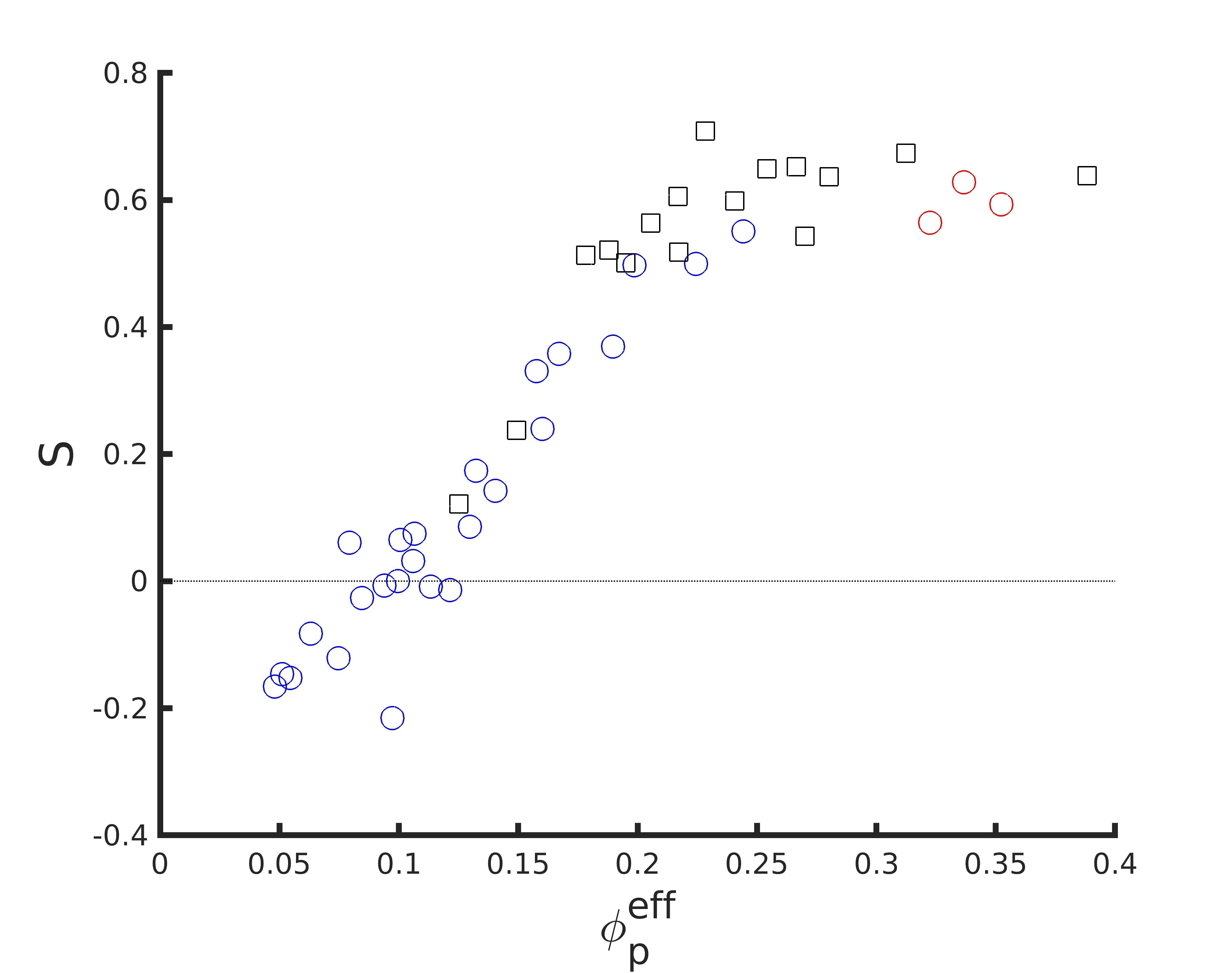}
\caption{Values of the orientational order parameter $S$ for different values of the effective volume fraction of ellipsoids $\phi_p^{eff}$ as of eq. \ref{eq:ellipse.phipeff}. Squares, blue circles and red circles stand for lamellar, cylindrical phase and inverted cylindrical phase, respectively.}
\label{fig:ellipse.phipeff}
\end{figure}

The orientation of the ellipsoids relative to the local interface is tracked by using the order parameter $S$ defined in equation \ref{eq:ellipse.S}. 
In Figure \ref{fig:ellipse.phase_diagram} a black dot is added for phase points in which $S>0.3$, that is, where orientational order is considerably high. 
Furthermore, we can define an effective filling fraction on the basis of A-compatible colloids having a reduced volume $V_A=f_0 V_{total}$ available space to occupy.
This effective filling fraction is \cite{huh_thermodynamic_2000}
\begin{equation}
\phi_p^{eff}=\frac{\phi_p}{f_{eff}}=\frac{\phi_p}{\phi_p +(1-\phi_p)f_0}
\label{eq:ellipse.phipeff}
\end{equation}
A plot of the orientational order parameter $S$ against the defined effective filling fraction of ellipsoids is shown in Figure \ref{fig:ellipse.phipeff} where disorder ($S \sim 0$) is found for low effective filling fraction. 
A rapid change in $S$ occurs as a moderate effective filling fraction is reached, while at the same time the cylinders-to-lamellae transition is induced. 
This suggests that the orientational order  strongly depends on the filling fraction of ellipsoids relative to the hosting domain, that is, ellipsoids need to be considerably constrained within their hosting domains. 
It is noted that inverted cylindrical phase (ie, ellipsoids occupying the majority of the space) display slightly lower order than lamellar ones, despite being at higher effective filling fraction. 
This is due to  higher local curvature of the interfaces that is characteristic of BCP cylindrical phases. 
This hypothesis is corroborated by analysing the snapshots in detail in the next figures. 

\begin{figure}[hbtp]
\centering
\includegraphics[width=0.9\linewidth]{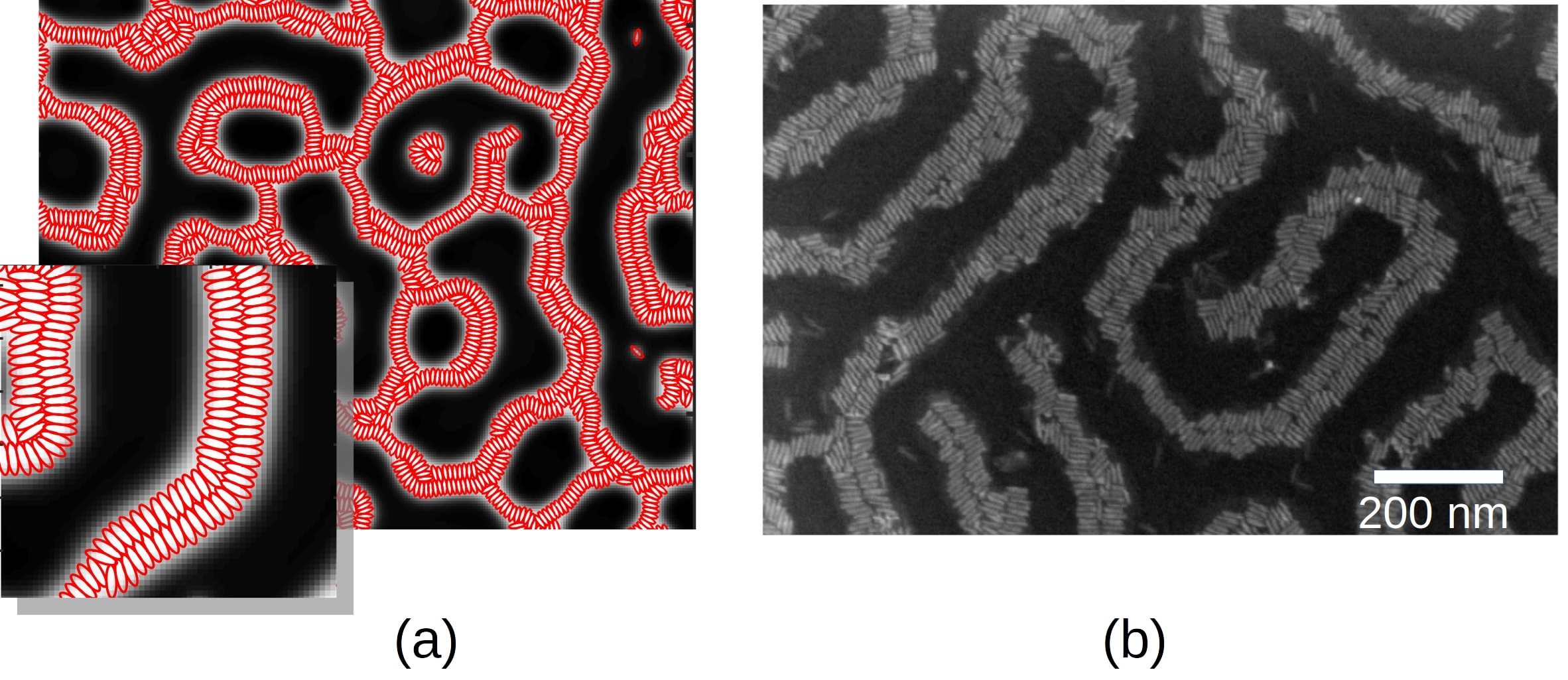}
\caption{
Moderate filling fraction of anisotropic NPs in diblock copolymer mixture. Comparison between  (a) simulations with inset and
 (b)  SEM image showing $33$ nm-long CdSe NRs co-assembled with PS-\textit{b}-PMMA with $H_0=132$ nm periodicity (PS domain size is $L_0=75$ nm). 
 The experimental NR diameter is $4.6$ nm with a filling fraction $0.26$.  
}
\label{fig:ellipse.comparison-highphip}
\end{figure}

Figure \ref{fig:ellipse.comparison-highphip} (a) shows an instance of ordering at a moderate filling fraction $\phi_p = 0.16$ and $f_0=0.4$ in a $256 \times 256$  grid system. 
One can notice that the side-to-side configuration is not homogeneous along the domains. 
Instead, we observe coexistence of both 1 and 2 rows, along with disordered states and even parallel (along the interface) orientation. 
Nonetheless, this behaviour appears more often near curved interfaces, as well as near defects of the lamellar structure. 
 These features can be found in experiments involving CdSe NRs  mixed with PS-\textit{b}-PMMA at a filling fraction $\phi = 0.26$.   
 This high resolution SEM image displays the side-to-side configuration of NRs within the PS domains.  
In Fig. \ref{fig:ellipse.comparison-highphip} (b),  1 and 2 rows occur for the same NR size and instances of disorder of parallel configuration appear, specially at the end of domains or at intersections, that is, defects in the lamellar structure.
Details of the experimental setup and initial conditions can be found in reference \cite{ploshnik_hierarchical_2010}. 

\subsection{Relative size of hosting domain/nanoparticle}

\begin{figure}
\centering 
\includegraphics[width=1.0\linewidth]{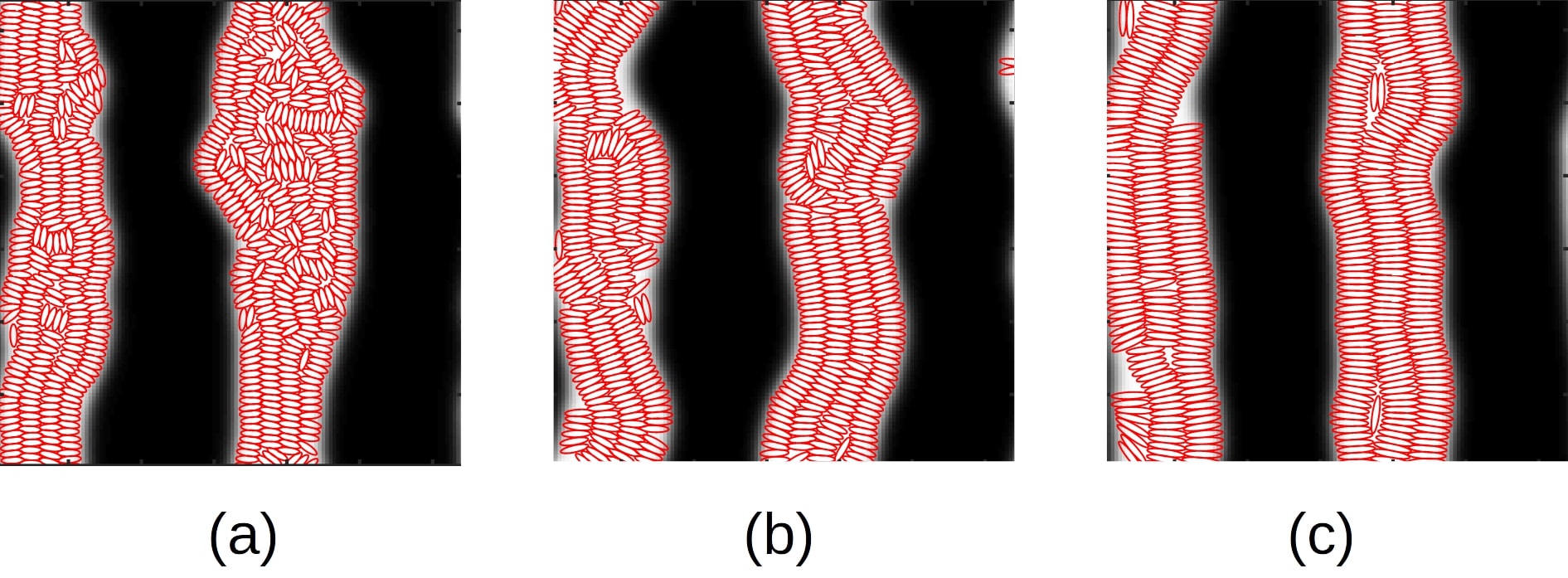}
\caption{
Final step of a system of ellipsoids with $b=0.6$ minor axis and three values of major  semi-axis
$a_0=2$, $2.5$ and $3.33$ for (a) (b) and (c), respectively. 
. The BCP constant $B=0.0002$ is used to result in a large lamella domain.
}
\label{fig:ellipse.rows}
\end{figure}

In Figure \ref{fig:ellipse.comparison-highphip} (b) the size of the NR major axis is chosen to fit two rows into a BCP lamellar domain. 
Similarly, in Figure \ref{fig:ellipse.comparison-highphip}  experiments and simulations show coexistence of 1 and 2 rows of anisotropic NPs. 
The role of the relative $2a/(H_0/2)$ size can be explored for a higher number of rows by simulating a larger periodicity, given by the parameter $B$ in the Ohta-Kawasaki free energy, which determines the value of the BCP periodicity $H_0$.  
Figure \ref{fig:ellipse.rows} shows simulations of 3 different sizes. 
The values are chosen to fit $3$, $4$ and $5$  rows. 
While the larger sizes in Figure \ref{fig:ellipse.rows} (b) and (c) show a well ordered configuration, the global order of the $a=2$ case is low (a). 
This is in accordance with experiments \cite{ploshnik_co-assembly_2010} in which smaller NRs displayed lower ordering.  

\begin{figure}[hbtp]
\centering
\includegraphics[width=0.75\linewidth]{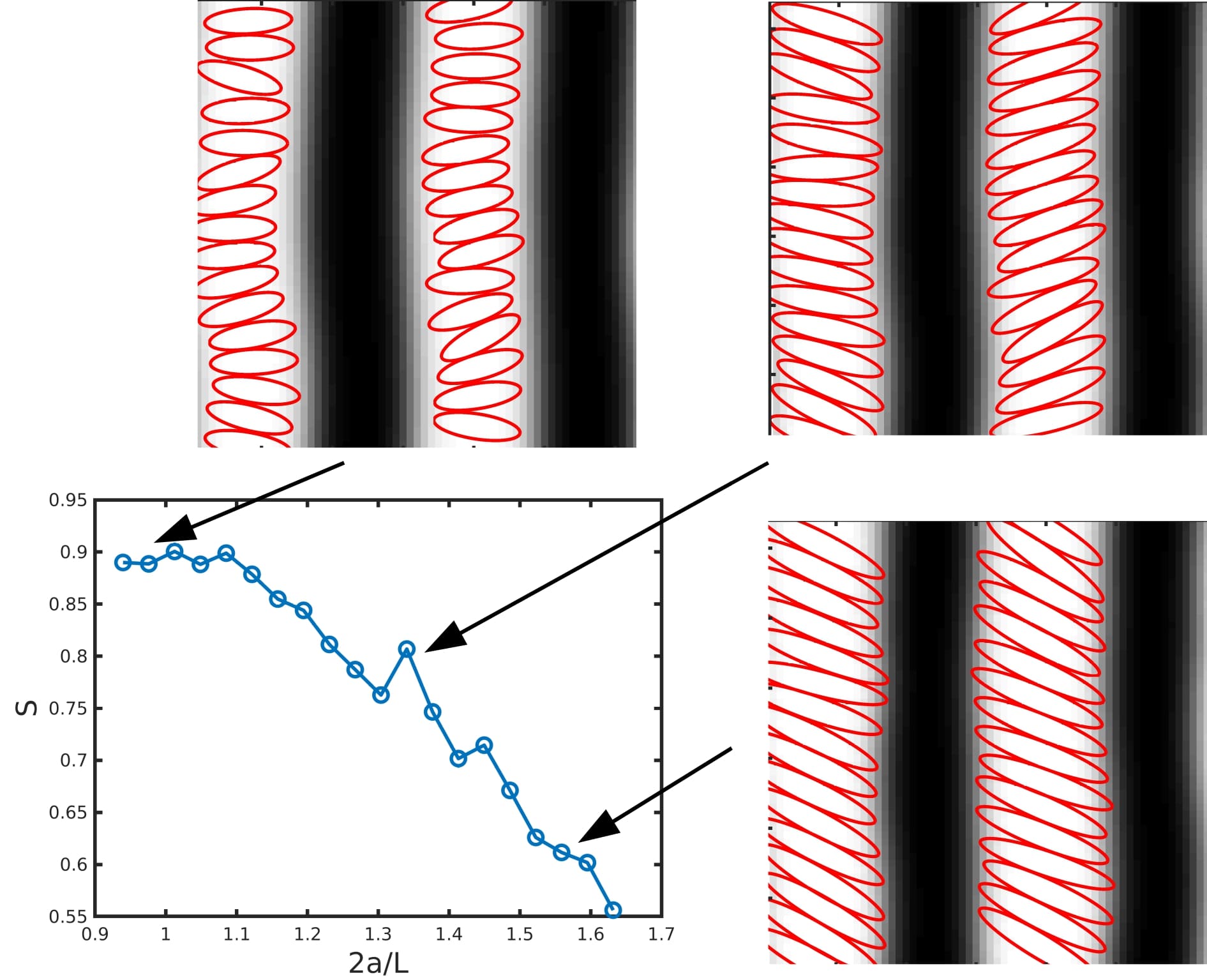}
\caption{Decrease of orientational ordering of ellipsoids when the size of the major axis $2a$ is larger than the available normal spacing $H_0/2-2\xi$ with $H_0$ the lamella periodicity and $\xi$ the interface semi-size}
\label{fig:ellipse.tilted}
\end{figure}

Composto et al \cite{deshmukh_two-dimensional_2007} showed that when the major dimension of NRs is larger than the lamella domain width, colloids tend to orient along the domain axis. 
While experiments have shown that smaller NRs orient normal to the domain direction, intermediate sizes can be explored using simulations. 
In Figure \ref{fig:ellipse.tilted} we explored the role of the ratio $2a/L$ with $2a$ being the effective major length of the ellipsoid and $L=H_0/2-2\xi$, which is a measure of the available horizontal spacing for NPs. 
We should note that inspection of the $\psi$ profile shows a clear weak segregation regime  for the BCP, which makes difficult to delimiter an interface/bulk region.  
In any case, the curve of $S$ along with the snapshots in Figure \ref{fig:ellipse.tilted} clearly shows a $S\sim 1$ regime when the ellipsoids can easily fit into the domains and normal to the interface. 
As the size of the NPs is increased, a slight rotation appears, which results in a decrease in $S$. 
In conclusion, we observe a tilted configuration when the
ratio between the major dimension of the ellipsoid and the BCP spacing is slighly larger than $1$.

\subsection{Low volume fraction of nanoparticles}

\begin{figure}[hbtp]
\centering
\includegraphics[width=0.9\linewidth]{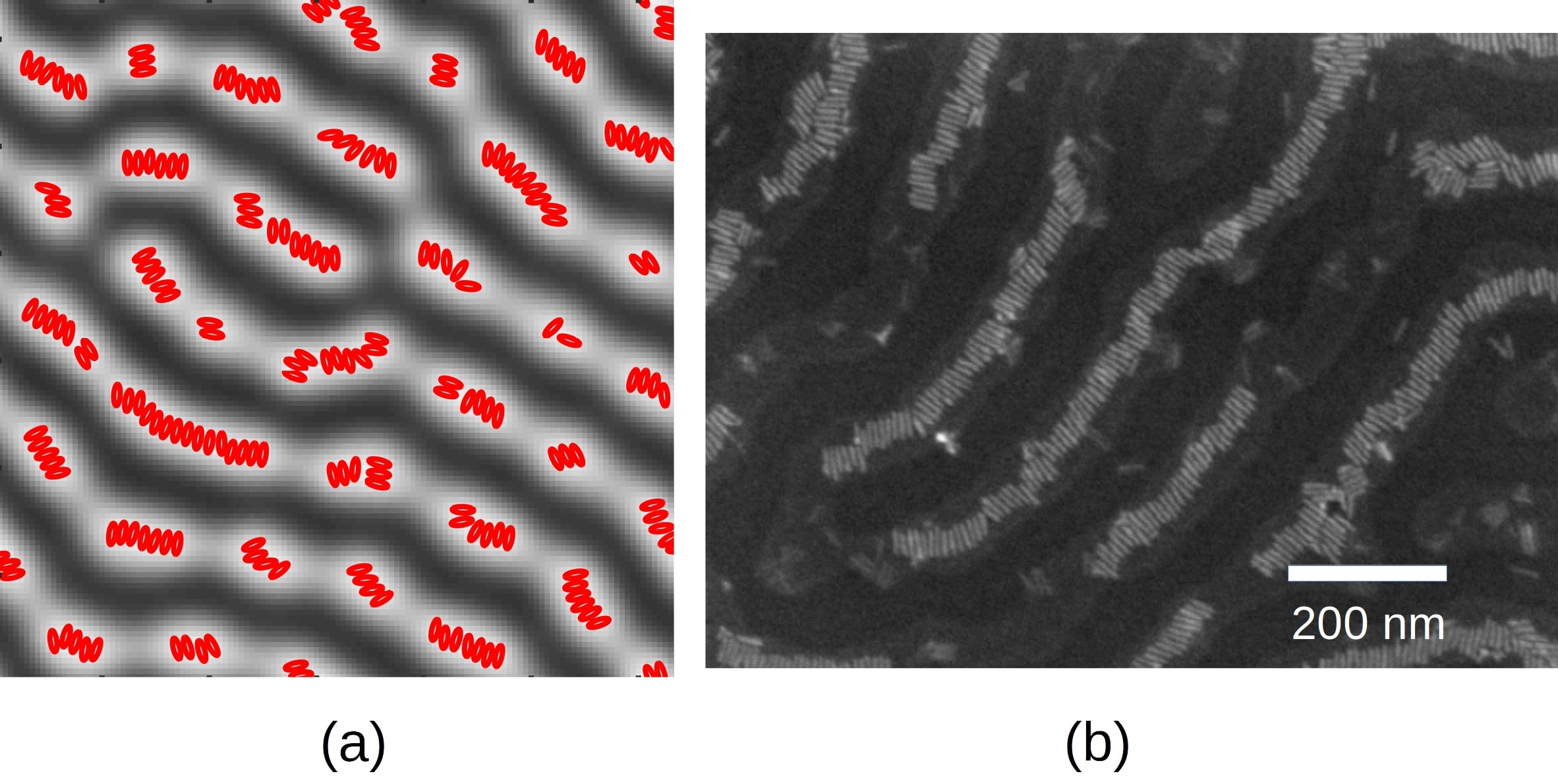}
\caption{Low filling fraction of anisotropic NPs in BCP, comparing (a) simulations and (b) SEM image of $33$ nm long CdSe NRs at a $0.15$ filling fraction. The diameter of the experimental NRs is $4.6$ nm. 
Unoccupied lamellar domain which is available for NPs is shown in light gray/white in (a) and as gray areas in (b).  
}
\label{fig:ellipse.lowphip}
\end{figure}

The side-to-side, perpendicular-to-domain-axis colloid configuration is shown to appear when the anisotropic NP occupies most of the hosting domain, thus the surrounding B-block boundary inflects a pressure. 
In Figure \ref{fig:ellipse.lowphip} we show that even at low filling fraction the normal configuration holds. 
This is due to the attractive interaction between colloids, which minimizes the free energy even at relatively low filing fractions. 
The simulations (a) show a resemblance with the experiments in (b), where NPs indeed form aggregates within their preferred domain. Defects (perpendicular and disordered orientation) are present both in simulations and experiments. 

\subsection{Role of the initial condition}

The hierarchical co-assembly displayed by NRs in experiments  depends on the initial arrangement of particles within the BCP before annealing\cite{ploshnik_co-assembly_2010,ploshnik_hierarchical_2010}. 
In particular, the  BCP was unable to break already-formed NP clusters. 
For that reason, simulations can be used to study the co-assembly starting from different initial conditions.
A competition between the tendency of attractive NPs to form aggregates on the one hand, and the equilibrium periodic morphology of the BCP on the other hand needs to be studied in detail. 
For that reason  two limiting regimes are selected: weakly and strongly interacting NPs with $U_0=0.001$ and $1.0$, respectively.

\subsubsection*{Weakly interacting nanoparticles}

\begin{figure}[hbtp]
\centering
\includegraphics[width=0.8\linewidth]{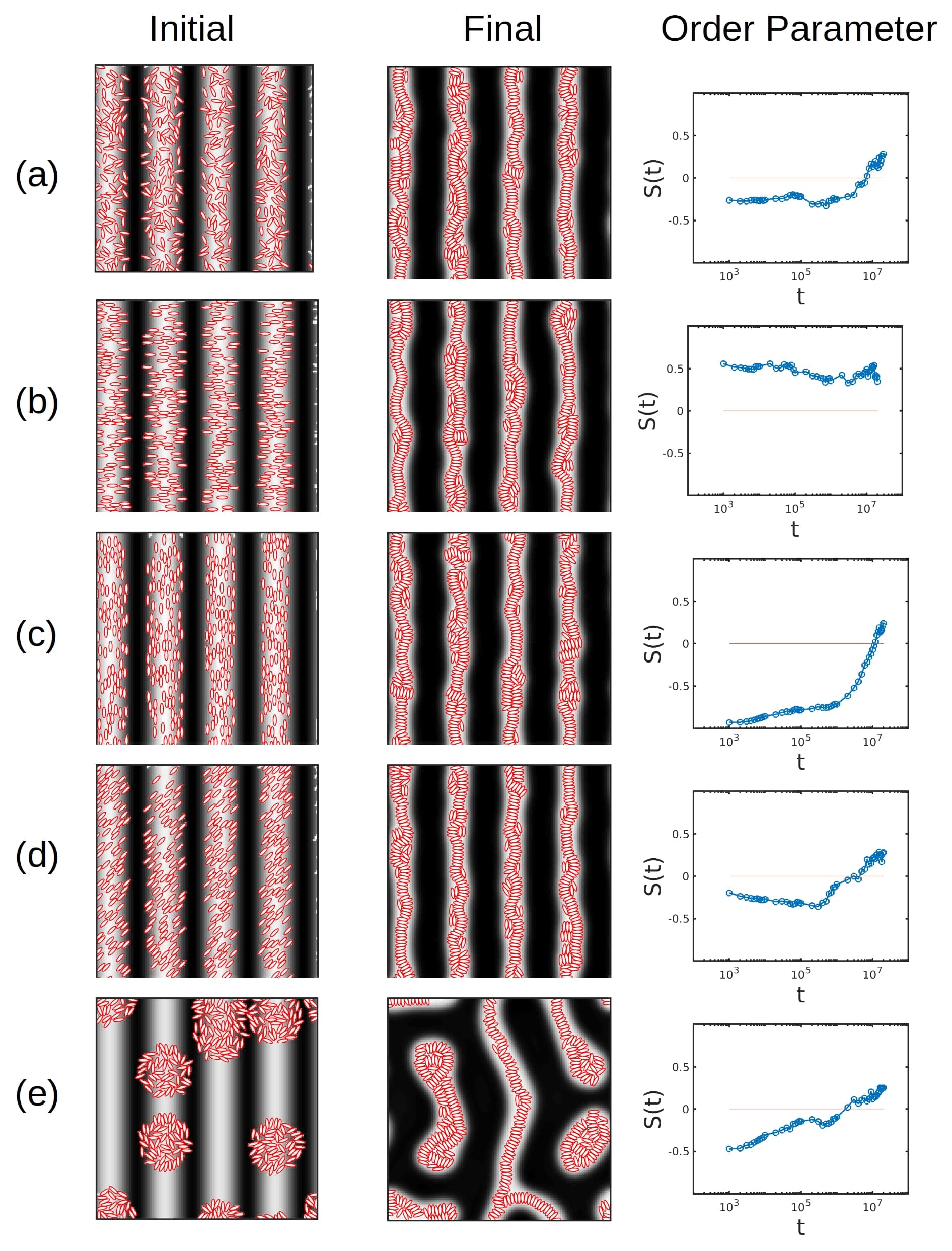}
\caption{Initial and final snapshot of several systems with different initial conditions. 
The time evolution of the orientational order parameter $S(t)$ is also shown for each case. 
In all cases the BCP is initially set to a sinusoidal concentration profile. 
In (a), NPs are randomly oriented and  positioned, within the white domain. 
In (b),(c) and (d), positions are again randomly chosen, while the orientation is $\phi_i=0,\pi/2,\pi/4$, respectively. In (e), the ellipses are placed randomly within clusters. }
\label{fig:ellipse.initial1}
\end{figure}

Figure \ref{fig:ellipse.initial1} shows five instances of the evolution of a system of relatively short particles with respect to the lamellar spacing of the diblock copolymer. 
The initial and final states are shown, while the order parameter $S(t)$ plot over time can be found in the right-most column. 
In all cases a dotted line marks the horizontal line $S=0$, so that instances of ordering $S>0$ are clear. 
In this figure, the BCP concentration profile is initialised as a sinusoidal, therefore, it is initially ordered.  
In (a), the NPs are randomly oriented and placed within the white domains. 
The system is then evolved into a final, ordered configuration of side-to-side ellipses. 
The system also exhibits alternating one and two rows of ellipses.
 (b), (c) and (d) show three different initial conditions regarding the orientation of the ellipses at $t=0$, respectively, $\phi_i=0,\pi/2$ and $\pi/4$. 
 Regardless of the initial condition, the final configuration is similar, meaning that this configuration is energetically favourable.
The order parameter $S$ in (a), (c) and (d)  reaches a final (approximately steady) state only at the very long stages of the simulation. 
Instead, the already-horizontal ordering of the  ellipses in (b) barely changes $S$ over time.  
Finally, (e) shows ellipses which are initially forming clusters in their preferred BCP domains. 
The final BCP morphology lacks the global orientation of the previous instances, since the NPs are initially forming clusters. Nonetheless, the orientation and ordering of ellipses is equally normal to the interface, which can be checked visually and by the positive $S$ value of the orientational order parameter.

\begin{figure}[hbtp]
\centering
\includegraphics[width=0.8\linewidth]{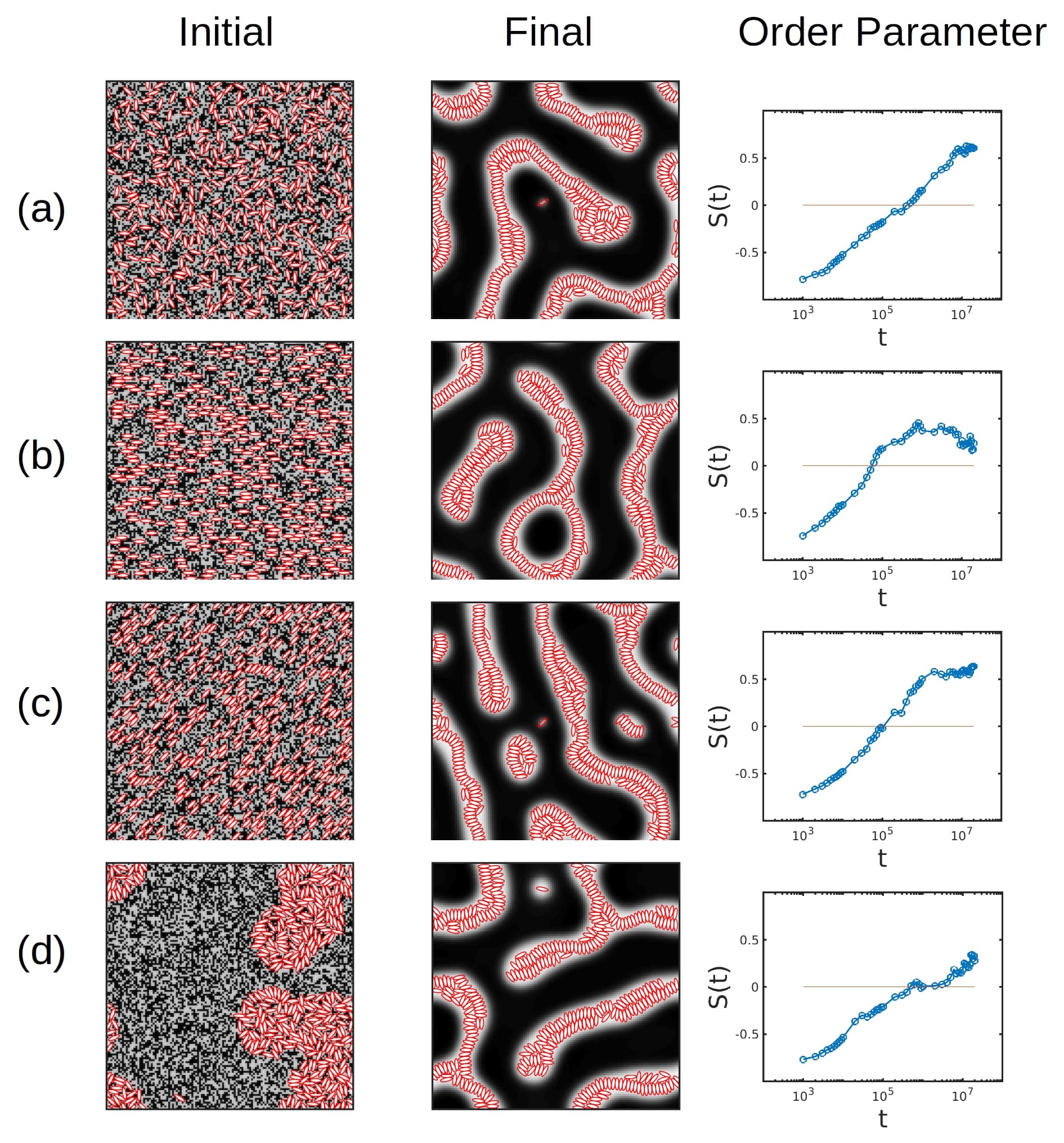}
\caption{Initial and final snapshot of several systems with different initial conditions. The time evolution of the orientational order parameter $S(t)$ is also shown for each case. In all cases the BCP is initially random(disordered). In (a) the orientation and position of all particles is chosen randomly. In (b) and (c), the position is chosen randomly, while the orientation with respect to the horizontal axis is $\phi_i=0,0.35\pi$, respectively. In (d) the particles are initially forming clusters. }
\label{fig:ellipse.initial2}
\end{figure}

Figure \ref{fig:ellipse.initial2} shows four instances of NR initial configuration in an initially disordered BCP. 
While the position of the particles is randomly chosen, the orientation is random for (a) and $\phi_i=0,0.35\pi$ for (b) and (c), respectively. The final state is shown in the central column whereas the evolution of the order parameter is displayed in the right column. 
In (d) the NPs are initially forming clusters without a collective orientation (disordered). 
In all of these cases the final state is a side-to-side configuration with ellipses oriented normal to the interface, that is, $S> 0$. 
Observing the evolution of $S(t)$ one can notice that the (b) and (c) cases reach the final $S$ value at a much shorter timescale than angularly-disordered cases (a) and (d).

\begin{figure}[hbtp]
\centering
\includegraphics[width=0.8\linewidth]{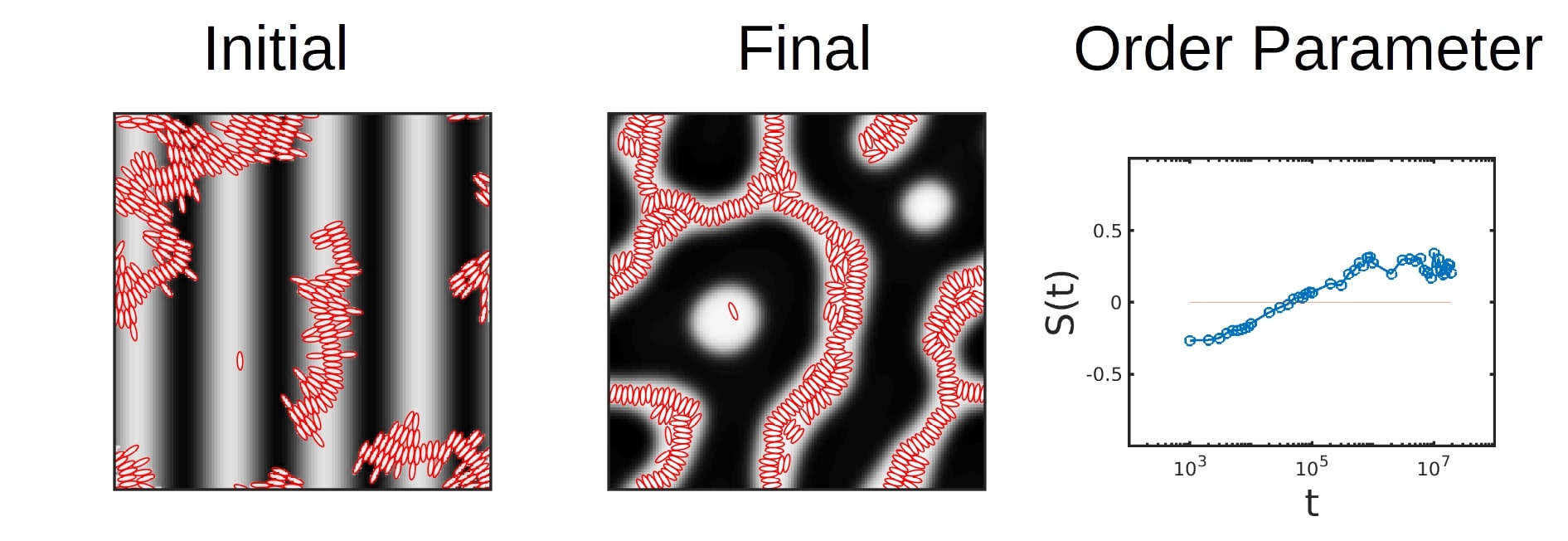}
\caption{Initially ordered  BCP with NP forming ordered clusters  at $t=0$. Initial and final snapshots are shown in the left and center figures while the order parameter $S(t)$ is displayed in the right-most figure.  }
\label{fig:ellipse.initial3}
\end{figure}

Figure \ref{fig:ellipse.initial3}   shows an initially ordered diblock copolymer, with NPs forming clusters with a low internal order (contrary to Figure \ref{fig:ellipse.initial1} (e) ) in which the internal orientation was random) . 
The time evolution suggests that the NPs are dispersed within the BCP, which is modified to accommodate the existing long-ordered sequences of ellipses. 
NPs form ordered arrays of side-to-side orientation within the white domains.

\subsubsection*{Strongly interacting nanoparticles}

In all of the above described cases, the BCP is able to acquire a stripe-like morphology, while NPs tend to appear relatively dispersed within the A domains, regardless of the initial condition.
 This was valid for particles which interact weakly between each other, such as  metal NPs that lack a fixed dipole moment. 
 Nonetheless, semiconductor NRs, such as CdSe, exhibit a  dipole moment which gives rise to a strong interparticle attractive interaction that leads to a strong tendencty towards particle aggregation\cite{ploshnik_hierarchical_2010}. 

\begin{figure}[hbtp]
\centering
\includegraphics[width=0.8\linewidth]{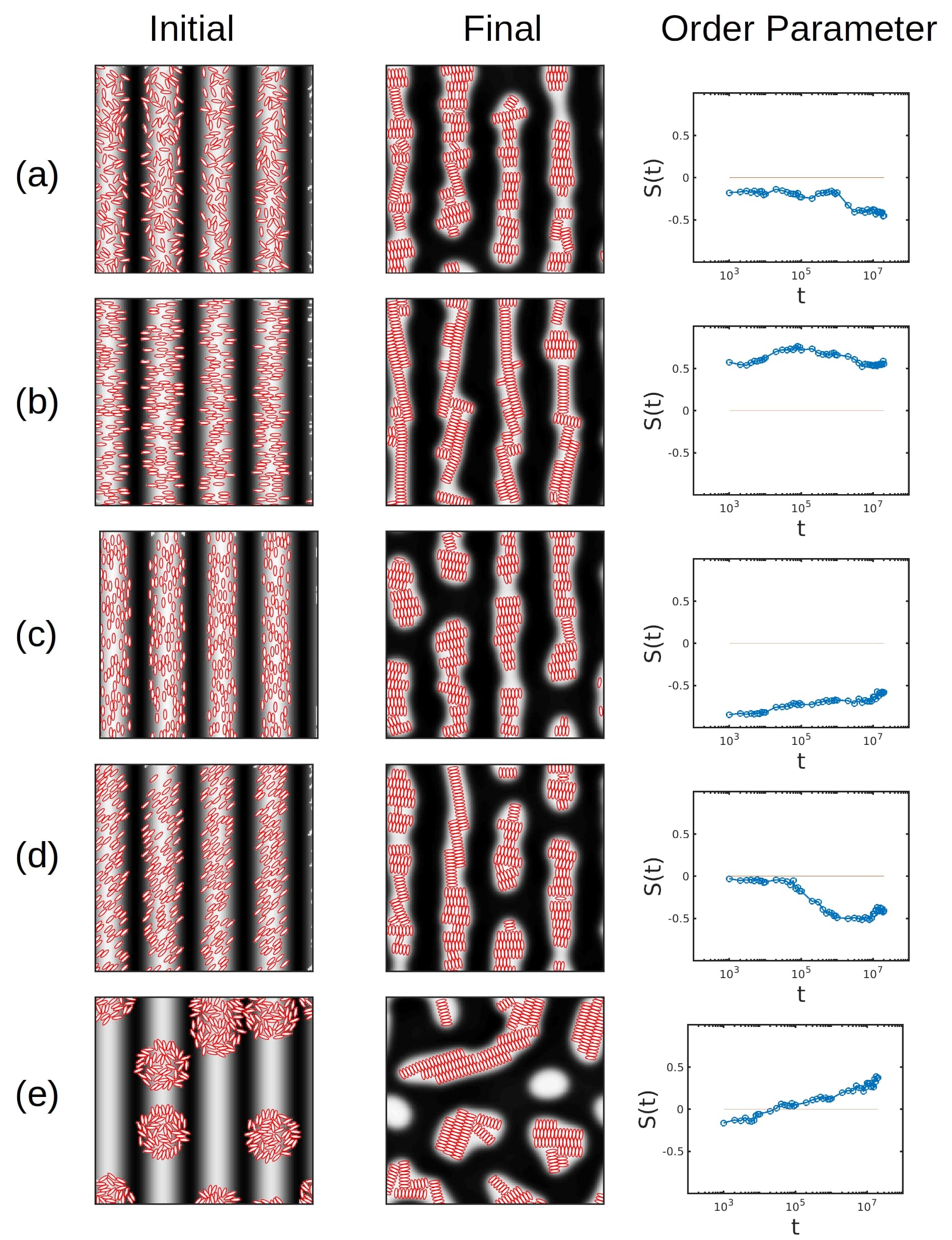}
\caption{Initial and final snapshot with different initial conditions (same as Figure \ref{fig:ellipse.initial1}) in the strong interaction regime $U_0=1$. The time evolution of the orientational order parameter $S(t)$ is also shown for each case. 
In all cases the BCP is initially fixed to a sinusoidal concentration profile.
 In (a), NPs have initial random orientation and  their position is random and confined to the white domains. 
 In (b),(c) and (d), positions are again randomly chosen, while the orientation is $\phi_i=0,\pi/2,\pi/4$, respectively. In (e), the ellipses are placed randomly within small clusters}
\label{fig:ellipse.initial_u0_1}
\end{figure}

\begin{figure}[hbtp]
\centering
\includegraphics[width=0.8\linewidth]{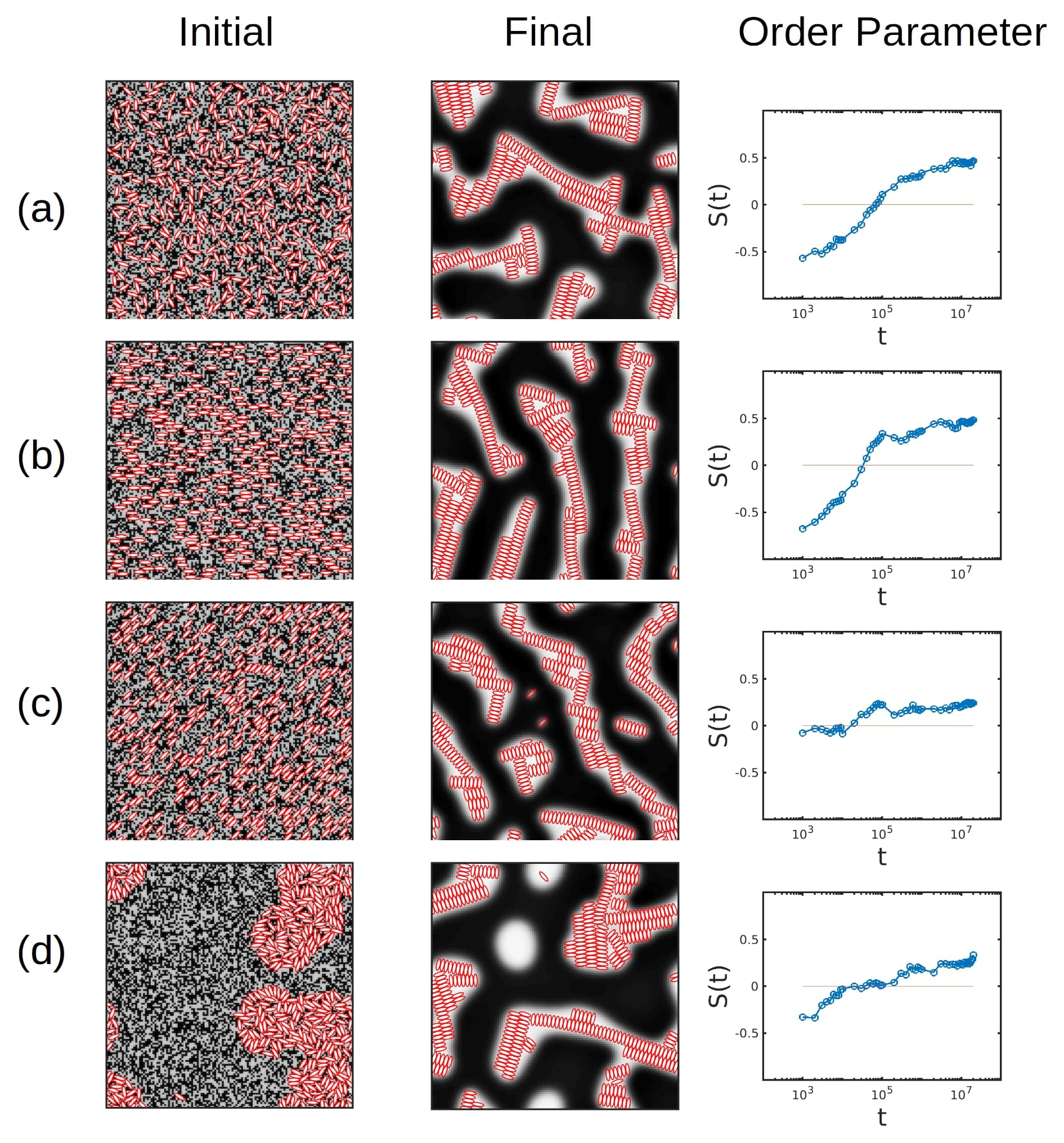}
\caption{Initial and final snapshot with different initial conditions (same as Figure \ref{fig:ellipse.initial2}) in the strong interaction regime $U_0=1$. The time evolution of the orientational order parameter $S(t)$ is also shown for each case. 
In all cases the BCP is initially random(disordered). In (a) the orientations and positions of all particles are chosen randomly. In (b) and (c), the position is chosen randomly, while the orientation with respect to the horizontal axis is $\phi_i=0 $ and $0.35\pi$, respectively. In (d) the particles are initially forming clusters. }
\label{fig:ellipse.initial_u0_2}
\end{figure}

\begin{figure}[hbtp]
\centering
\includegraphics[width=0.8\linewidth]{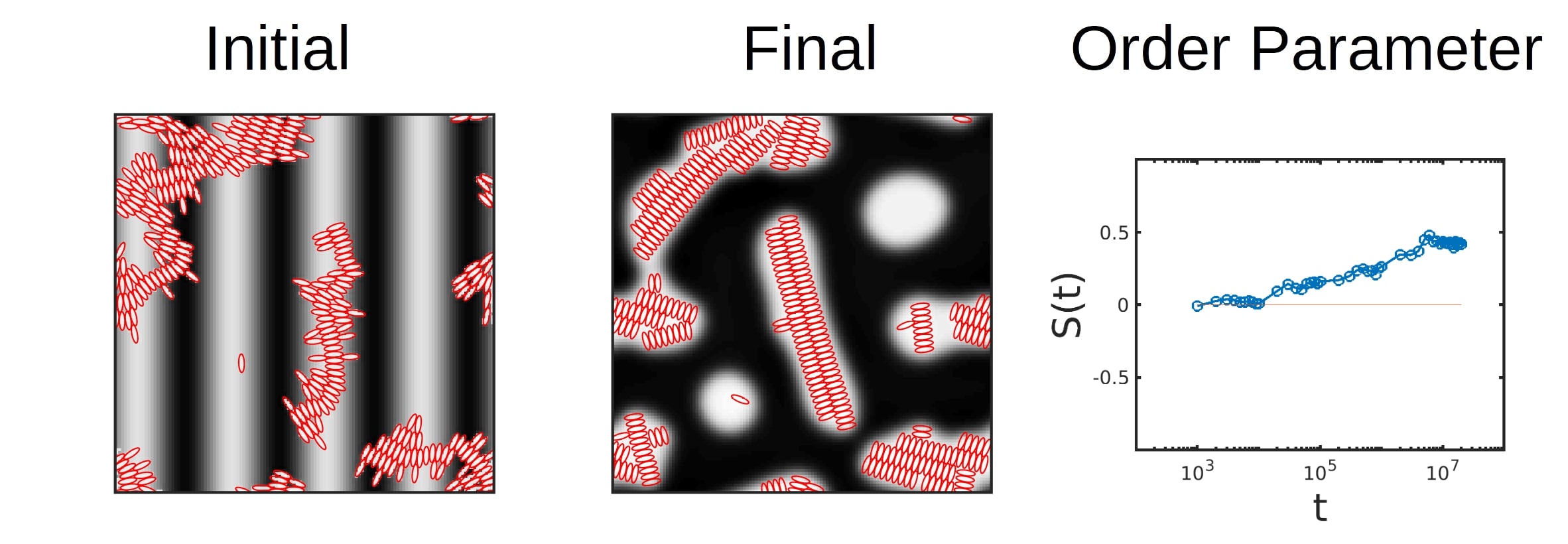}
\caption{
Initially ordered  BCP with NP forming ordered clusters  at $t=0$. Initial and final snapshots are shown in the left and center figures while the order parameter $S(t)$ is displayed in the right-most figure. The NP-NP interaction parameter is $U_0=1$.
}
\label{fig:ellipse.initial_u0_3}
\end{figure}

A strong interparticle potential scale parameter $U_0=1$ can be used to understand the co-assembly behavior in the strong interparticle potential limit, as shown in figures \ref{fig:ellipse.initial_u0_1},\ref{fig:ellipse.initial_u0_2} and \ref{fig:ellipse.initial_u0_3}. These are directly related to figures \ref{fig:ellipse.initial1}, \ref{fig:ellipse.initial2} and \ref{fig:ellipse.initial3} by selecting the same initial conditions. 

Comparing figures \ref{fig:ellipse.initial1} (weakly interacting) and \ref{fig:ellipse.initial_u0_1} (strongly interacting) we can clearly draw the conclusion that at $U_0=1$ the NPs are driving the co-asembly, with the BCP domains being formed around aggregates of colloids (except for (b), as compared with the weakly interacting case, in which the BCP tended to form elongated lamellar-like domains. Ordering in the NPs is also strongly dominated by the initial configuration in the $U_0=1$ regime, with $S$ reaching positive values only in (b) and (e) cases. 
A comparison between the curve of $S(t)$ in Figure  \ref{fig:ellipse.initial1} (c) and \ref{fig:ellipse.initial_u0_1} (c)  leads to the conclusion that in the weakly interacting NPs regime the NPs undergo first a dispersion within the BCP domains, while in a slower time scale the NPs achieve global normal orientation with the BCP interface. 
This global ordering is not present at $U_0=1$, where the strong interparticle potential rapidly assembles the NPs into vertically oriented groups of particles in aggregates. Similarly, in (e) the aggregated particles at $t=0$ tend to form clusters also after the time evolution, with the BCP clearly being forced to form less elongated domains.

The initial orientation of NPs when the BCP is initially disordered also affects the co-assembly in the strong interacting NP regime. 
In Figure \ref{fig:ellipse.initial_u0_2}, compared to its equivalent shown in figure \ref{fig:ellipse.initial2}, clearly displays less elongated domains, with NPs more prone to form well-ordered structures, while at the same time forming more aggregates that enhance the local size of the hosting domains. 
Again, this suggests that the BCP is unable to complete its assembly by disassembling the aggregates, instead, it merely forms domains around aggregates of ellipses.

Similarly, in Figure \ref{fig:ellipse.initial_u0_3}   the initial NP aggregates cannot be broken into elongated domains to the same degree as occurred in Figure \ref{fig:ellipse.initial3}. Instead, the BCP forms domains around clusters of particles. Since the initial aggregates were already made of considerably well-ordered NPs, the final $S$ value is particularly high, meaning that a high ordering is achieved.

In summary, the role of the initial condition is crucial when the NPs interact strongly with each other. This strong interaction leads to the formation of aggregates in the early stages of the time evolution of the system that are not broken by the BCP evolution (whether it is from disorder to order, or an already phase-separated BCP). Weakly interacting NPs, on the other hand, undergo a co-assembly on a similar time scale as the BCP, therefore,  the side-to-side along with the normal-to-interface configuration is easily obtained under any initial condition.

\subsection{Role of energy parameters}

Simulations can be used to gain insight on the effect that the interaction parameters have on the formation of the side-to-side configuration normal to the interface between domains. 
A simple energy analysis for NRs suggests that this configuration is energetically preferential both for the inter-colloidal potential and the NP-polymer coupling \cite{ploshnik_co-assembly_2010,ploshnik_hierarchical_2010}. 

\begin{figure}[hbtp]
\centering
\includegraphics[width=.75\linewidth]{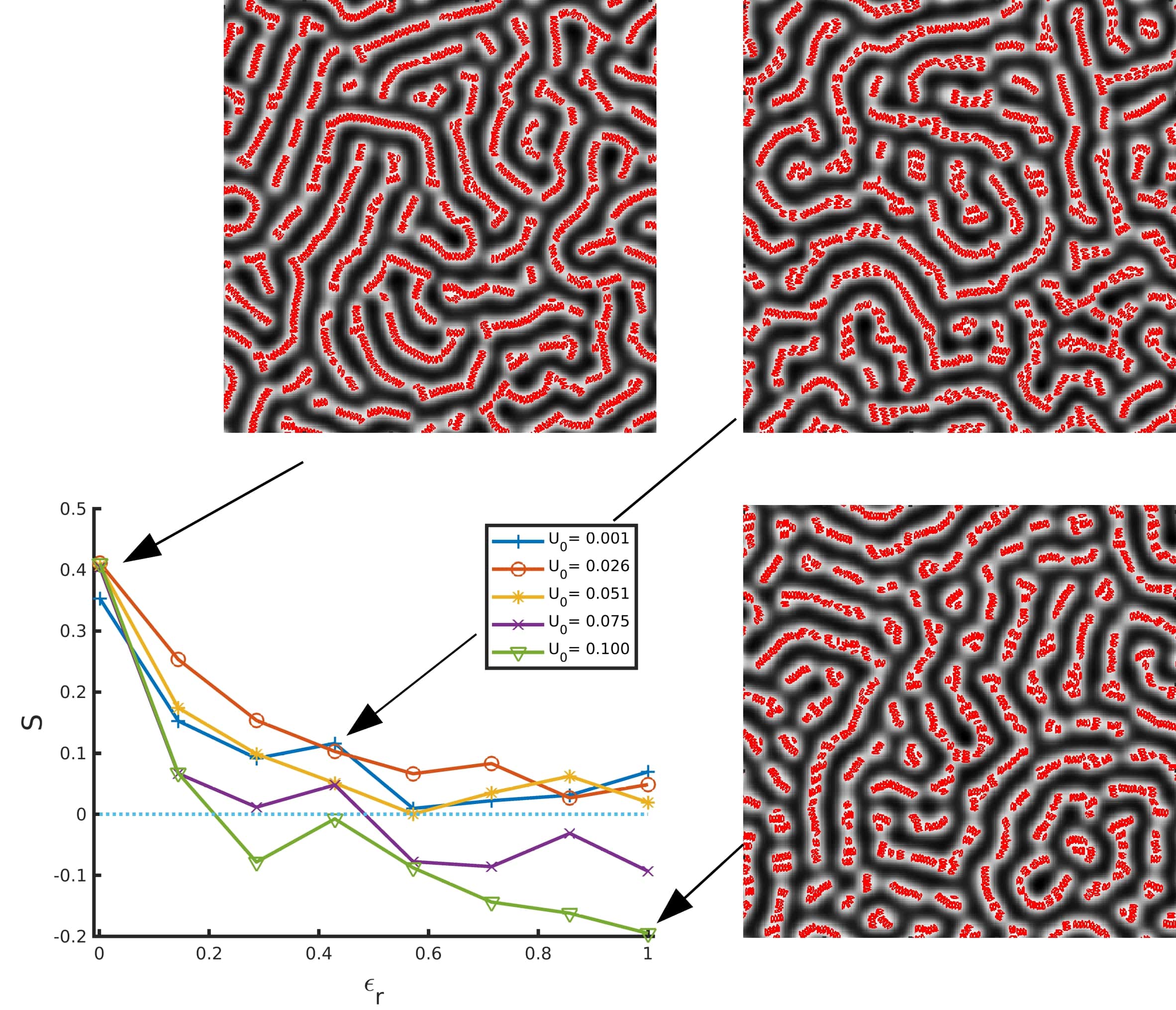}
\caption{Orientational order parameter $S$ for different final stages tuning the interparticle potential parameters: $U_0 $ and $\epsilon_r$, the strength and anisotropy of the Gay-Berne potential, respectively. 
Three snapshots of the representative parameters are shown. }
\label{fig:ellipse.epsr}
\end{figure}

The interparticle Gay-Berne potential described in the Model section sets the interaction between two ellipses with two energetic parameters: $U_0$ sets the scale of the interaction, while $\epsilon_r$ describes the anisotropy in the depth of the potential minima. For that reason, in Figure \ref{fig:ellipse.epsr} we explore these two parameters via the orientational order parameter $S$. It is clear that the anisotropy value $\epsilon_r$ is key on the formation of the side-to-side configuration as we find $S\sim 0$ as the anisotropy of the potential is closer to $1$. This leads to more tip-to-tip configurations, which in turn are better accommodated with the NPs oriented along the domains, as can be found in the two snapshots in the right hand-side of Figure \ref{fig:ellipse.epsr}. Contrary to that, high anisotropy leads to well-ordered side-to-side configuration. Because the lamella domain spacing is  similar to the major size of the ellipses, the BCP accommodates only one row of ellipses.  

Figure \ref{fig:ellipse.epsr} shows that the energetic scale $U_0$ plays a less relevant role than the anisotropy factor $\epsilon_r$. 
We observe that even at low values of $U_0$ the ordering is kept $S>0.4$ which is indicative that the configuration is stable even at low interparticle potential strengths. 
These results are in accordance with a simple energy analysis shown in  the Supplementary Information, where we conclude that in order to have an energy minimum in the side-to-side configuration, $e>>\epsilon_r$  should  be satisfied. 

\begin{figure}[hbtp]
\centering
\includegraphics[width=0.75\linewidth]{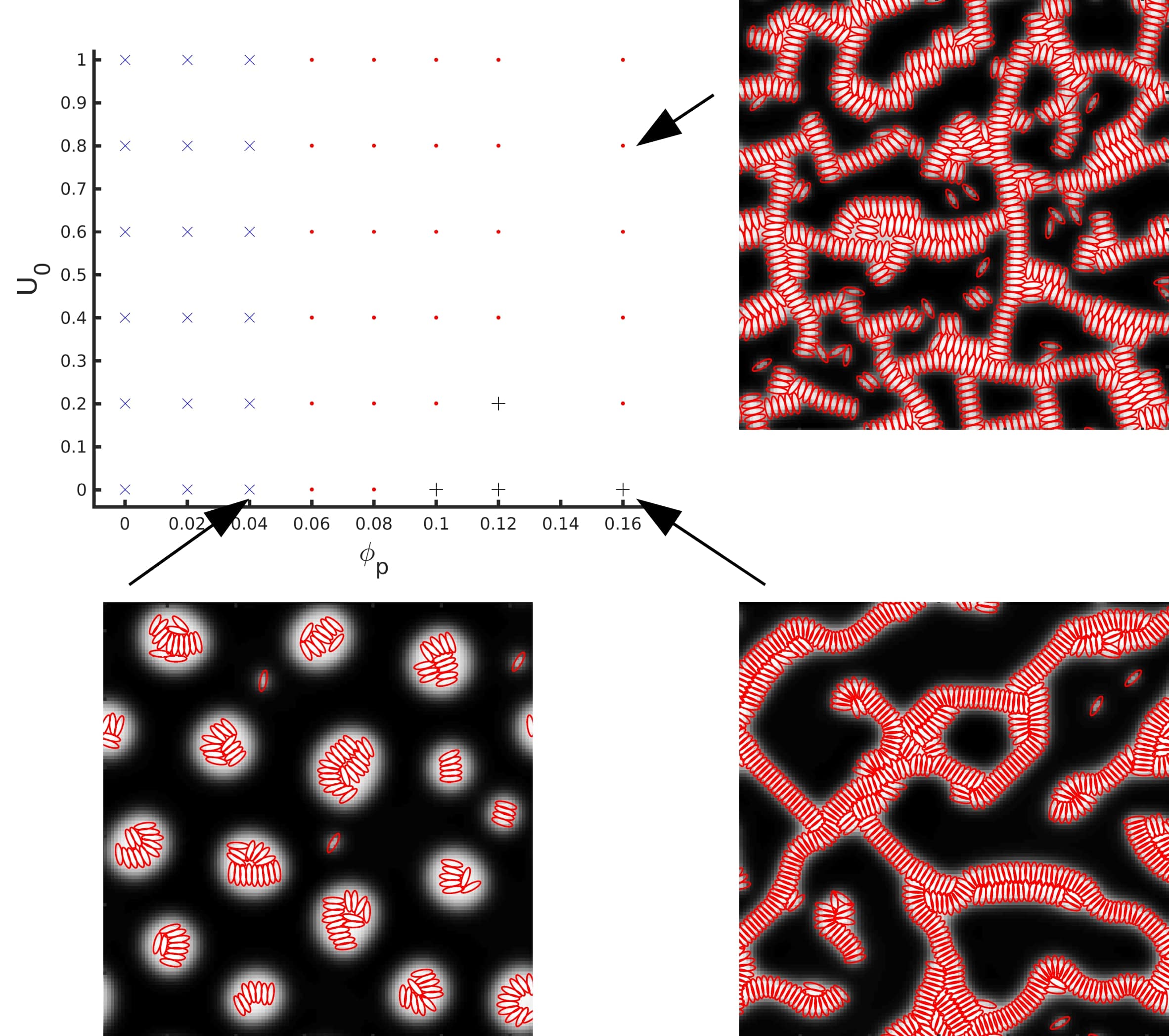}
\caption{Phase diagram of the assembly of ellipses in a diblock copolymer with $f_0=0.3$. The number of particles is explored in the X axis $\phi_p$ and the strength of the interparticle potential is tuned via $U_0$. Markers relate to the value of the orientational order parameter as: blue cross   {\color{blue} x} for $S<0.01$; red dots {\color{red} $\boldsymbol{\cdot}$} for $0.01<S<0.3$; and black plus sign {\color{black} +} for $S>0.3$ } 
\label{fig:ellipse.phd-U0-phip}
\end{figure}

Such energy analysis is considered only in the case of relatively high filling fraction of NPs in the system, a regime in which the particles-to-polymer coupling is considerably strong, as the contacts between the colloids and the interfaces become important. 
Figure  \ref{fig:ellipse.phd-U0-phip} shows a phase diagram of the filling fraction of particles in the system $\phi_p$, and the strength of the potential $U_0$. 
The cylinder-forming neat BCP is chosen by setting $f_0=0.3$. 
The orientational order is characterised  by the order parameter $S$.
In the low filling fraction regime we can find ellipsoids segregated within the minority white domains without a particular orientation with respect to the BCP interface.
On the other hand, at high filling fraction the nanoparticles can induce a transition into elongated BCP domains. 
In this regime, a lower value of  $U_0$  leads to a higher ordering. 
Contrary to that, a large interaction strength leads to the interparticle potential driving the ordering behaviour of the system. 
In this regime the NPs are less prone to minimise the contacts with the black domains, and minimisation of the interparticle potential is dominant enough. 
This result agrees with the conclusion drawn from the comparison between figures \ref{fig:ellipse.initial1}-\ref{fig:ellipse.initial3} 
and figures  \ref{fig:ellipse.initial_u0_1}-\ref{fig:ellipse.initial_u0_3} 
which showed that a lower interaction strength led to a higher degree of ordering. 

We can therefore conclude that at low volume fraction,  strong interparticle interaction is necessary for obtaining ordered NP superstructures.
 This is not the case at higher filling fraction, in which merely the NP-BCP interaction is enough to ensure that the NP will assemble in the described configuration.

\section{Conclusions}

The co-assembly of anisotropic nanoparticles in BCPs has been studied by means of mesoscopic simulations in the case of A-modified NPs with an  elliptical shape. 
Ellipsoidal nanoparticles have been shown to induce phase transitions in the  block copolymer matrix due to an increase in the effective concentration of the hosting copolymer. 
In turn, the combination of BCP-NP coupling and intercolloidal attractive forces leads to a well-ordered configuration of anisotropic colloids which reproduces experimental results \cite{ploshnik_co-assembly_2010,ploshnik_hierarchical_2010}. 
When confined within one of the BCP phases, ellipsoids are found to orient normal to the domain axis in order to minimise the contacts with the surrounding incompatible phase while minimising the angle-dependent intercolloidal potential. 
Ellipsoidal colloids are used to mimic CdSe nanorods mixed with PS-\textit{b}-PMMA used in experiments. 
Direct comparison between microscopy images and simulations shows  considerable similarity between simulations and experimental results. 
Despite the differences in the dynamic path between the experiments (solvent vapour annealing and three dimensional thickness of the film) and  the simulations, we show that the final state of the ordered configuration is reproducible with different initial conditions resulting in side-to-side nanorods. 
This suggests that the simulated model captures the most dominant factors in the co-assembly, that is, interactions between the components of the system and the filling fraction.

The size of the NP with respect to the BCP periodicity plays a crucial role in the assembly of anisotropic colloids. 
Smaller NPs tend to form more rows than large ones for a given BCP periodicity, whereas the ordering increases with larger particles, which is in accordance with experiments. 
Furthermore, NPs that are slightly larger than the lamellar spacing undergo a rotation with respect to the interface, while maintaining the side-to-side intercolloidal organisation. 

A study of several different initial conditions (both initially ordered and  disordered) has drawn the conclusion that weakly interacting NRs organise side-to-side within a phase-separated BCP that achieves a lamellar morphology, regardless of the initial condition. 
In this regime the BCP can undergo the usual phase-separation even in the case of an initially-clustered NP configuration. 
On the other hand, the initial configuration of colloids is crucial in the case of strongly-interacting nanoaparticles, which are trapped in a metastable state that does not allow the system to reach a side-to-side organisation. 
This occurs, for example, if NPs are initially forming aggregates, which the BCP is unable to break-up.  
Weakly interacting NR's behaviour can be related to metal NRs which lack a fixed dipole moment, in which the co-assembly is mostly dictated by the block copolymer morphology. 
Semiconductor NRs such a CdSe display a fixed dipole moment ($3.3 \times 10^{-28}$ C m for $33$ nm rods\cite{li_origin_2003}, for example) leading to side-to-side organisation even at low concentrations. 
Even higher NR-NR interaction such as ZnO NR ( dipole moment of $4.1 \times 10^{-26}$ C m for $33$ nm rods\cite{dag_large_2011}) can be related to a high value of $U_0$. 
Finally, a high energy anisotropy in the colloid-colloid potential has been shown to be crucial for determining the final side-to-side/normal to domain axis configuration.

In summary, we have presented a computational method that mimics a complex co-assembly process of anisotropic NPs in BCPs. 
We have been able to gain insight over the role of several parameters that are experimentally difficult to explore. 
We have identified the importance of the relative size between the NP main axis and the lamellar spacing, which dictates the relative orientation and the  number of rows in the assembly. 
Two different energy regimes have been identified: weakly and strongly interacting NPs undergo different types of co-assembly with the BCP, which corresponds to semiconductor and metallic NPs. 
Weakly-interacting NPs with a high energetic anisotropy have been shown to display the highest level of side-to-side configuration while allowing the BCP lamellar morphology to fully develop.

\begin{acknowledgement}
I. P. acknowledges support from MINECO (Grant No. PGC2018-098373-B-100), DURSI (Grant No. 2017 SGR 884) and SNF Project No. 200021-175719.
The authors thank Elina Ploshnik, Asaf Salant and Uri Banin for their contribution to the experimental results shown in the paper. 
Financial support was provided by the Israeli
Science Foundation, grant number 229/17.
JD thanks the BritishSpanish Society for financial support. 
\end{acknowledgement}

\begin{suppinfo}

A simplified energetic analysis to justify the side-to-side, normal to interface orientation of nanorods

\end{suppinfo}

\bibliography{references}

\end{document}